\documentclass[conference]{IEEEtran}
\usepackage[utf8]{inputenc}
\usepackage[english]{babel}
\usepackage[T1]{fontenc}
\usepackage{lmodern}
\usepackage{cite}
\usepackage{amsmath,amssymb,amsfonts}
\usepackage{algorithmic}
\usepackage{graphicx}
\usepackage{textcomp}
\usepackage{xcolor}
\usepackage{url}
\usepackage{booktabs}
\usepackage{microtype}
\usepackage{tabularx}
\usepackage{tikz}
\usetikzlibrary{shapes.geometric, arrows.meta, positioning, calc, shadows}

\normalsize

\begin{document}

\title{An Embedded Ergonomic Posture Monitoring and AI-Assisted Home Learning System on a Raspberry Pi 3B Edge Platform}

\author{
\IEEEauthorblockN{Truyen T. Le\IEEEauthorrefmark{1}, Vinh Dang Ngoc\IEEEauthorrefmark{1}, and Thai Bao Huynh\IEEEauthorrefmark{2}$^*$}
\IEEEauthorblockA{\IEEEauthorrefmark{1}\textit{Nguyen Hue High School}, Tuy Hoa, Phu Yen, Vietnam \\
Email: letrongtruyenofficial@gmail.com, dnvinh.thpt.nh@daklak.edu.vn}
\vspace{0.4em}
\IEEEauthorblockA{\IEEEauthorrefmark{2}\textit{University of Information Technology, Vietnam National University Ho Chi Minh City}, Ho Chi Minh City, Vietnam \\
Email: 23520105@gm.uit.edu.vn \\
$^*$Corresponding author}
}

\maketitle

\begin{abstract}
During independent home study sessions, students frequently sustain improper sitting postures, accumulate excessive uninterrupted sedentary hours, and experience productivity loss caused by off-task digital media distractions. Furthermore, learners commonly lack real-time pedagogical scaffolding when encountering demanding self-study curricula. To mitigate these interrelated ergonomic and instructional challenges, this paper presents an affordable, privacy-preserving cyber-physical embedded system designed specifically for domestic study spaces. Engineered on a single-board computer (Raspberry Pi 3 Model B, 1\,GB RAM) paired with a UVC optical camera and a USB condenser microphone, the system executes on-device computer vision utilizing lightweight landmark inference (MediaPipe Pose) and OpenCV to monitor upper-body spatial alignment, sagittal head inclination, and coronal shoulder asymmetry in real time without persisting raw video streams. A deterministic session management finite state machine (FSM) quantifies active study durations, enforces health-compliant rest periods, and detects illicit smartphone usage. Simultaneously, an integrated multimodal artificial intelligence tutoring agent parses voice and text queries via an encrypted API gateway, synthesizing Socratic conceptual hints rather than direct solutions. A lightweight parental web dashboard delivers encrypted session analytics and scheduling configurations. Empirical benchmarks executed on the physical prototype confirm an end-to-end inference latency of $244$--$250$\,ms at a $352\times 288$ analysis resolution (sustaining $4.0$--$4.1$ inference frames per second), a trunk-angle estimation error of $4.39^\circ$ mean absolute error over $5760$ synthetic ground-truth poses, a resident set size of $320$\,MB, a peak SoC temperature of $58.0^\circ$C under sustained load, and a total latency of $1.58$\,s for the offline TTS feedback path, proving high practicality for resource-constrained edge deployment.
\end{abstract}


\section{Introduction}

The rapid proliferation of digital pedagogical curricula, remote learning platforms, and self-directed academic environments has fundamentally shifted the locus of primary and secondary education into the home. While digital tools democratize access to academic information, independent study without continuous ergonomic and supervisory oversight introduces severe physical and cognitive vulnerabilities for adolescent students.

From a physiological perspective, prolonged seated screen interaction frequently induces musculoskeletal disorders. Sustained static flexion of the cervical spine---commonly termed ``text neck'' syndrome---magnifies compressive forces on cervical intervertebral discs, while slouched thoracic kyphosis and asymmetrical lateral pelvic loading induce chronic lower back strain and visual fatigue. Epidemiological assessments confirm that sedentary behavior exceeding uninterrupted 45-minute intervals impairs circulatory dynamics and degrades sustained cognitive attention. Concurrently, unmonitored digital home environments expose students to cognitive fragmentation caused by non-educational mobile devices, gaming applications, and social media notifications.

From a supervisory and academic standpoint, modern working parents cannot provide uninterrupted observation during after-school study sessions. Manual parental spot-checks often precipitate familial tension and fail to deliver objective empirical metrics regarding genuine study engagement, break compliance, and ergonomic posture health. Furthermore, when students confront complex mathematical derivations or scientific concepts independently, existing commercial search utilities and generative language models often output direct answers immediately. This direct answer delivery short-circuits cognitive struggle, cultivating academic dependency rather than reinforcing analytical problem-solving abilities.

To resolve these interconnected ergonomic, supervisory, and instructional challenges, this paper develops an integrated, low-cost cyber-physical system tailored for home learning environments. The core framework couples localized edge vision inference on an embedded single-board computer (Raspberry Pi 3 Model B) with an interactive, cloud-orchestrated multimodal artificial intelligence tutoring agent.

The primary contributions of this paper are summarized as follows:
\begin{itemize}
    \item \textbf{Privacy-Preserving Edge Vision Pipeline:} A lightweight, on-device pose estimation architecture that evaluates multi-planar ergonomic indices (sagittal trunk flexion, craniovertebral neck tilt, and coronal shoulder asymmetry) in real time while guaranteeing complete video privacy through volatile, in-memory processing.
    \item \textbf{Deterministic Ergonomic Session Manager:} A finite state machine (FSM) that automatically tracks net academic engagement, coordinates periodic rest breaks aligned with ergonomic health guidelines, and flags off-task mobile screen distractions.
    \item \textbf{Socratic Multimodal AI Tutoring Agent:} A hybrid conversational agent capable of ingesting speech and text, utilizing structured system prompts to provide step-by-step conceptual scaffolding rather than ready-made solutions, with automatic model-fallback resilience against cloud-side transient failures.
    \item \textbf{Comprehensive Embedded Benchmarking:} Extensive empirical validation on resource-limited ARM architecture profiling frame rates, geometric estimation accuracy, memory footprint, and thermal dissipation under sustained real-time load on the deployed prototype.
\end{itemize}

\section{Related Work}

\subsection{Ergonomic Posture Tracking Technologies}
Ergonomic monitoring systems traditionally relied on wearable sensors, including chest straps, electromyography (EMG) patches, and multi-axis Inertial Measurement Units (IMUs) affixed to the spine and cervical vertebrae. While IMUs yield precise kinematic angle measurements, wearable hardware introduces significant practical limitations for school-aged children: physical discomfort, skin irritation, recurring calibration drift, and the persistent burden of battery recharging.

Consequently, vision-based non-intrusive posture estimation has garnered widespread research interest. Deep convolutional neural network (CNN) architectures such as OpenPose deliver robust multi-person joint coordinates but impose computational demands exceeding the hardware capabilities of standalone edge microprocessors. The introduction of optimized two-stage regression pipelines---notably Google MediaPipe Pose \cite{mediapipe}---enables single-person landmark inference on commodity CPU threads. However, deploying such pipelines on embedded processors like the Broadcom BCM2837 requires strict resolution optimization, algorithmic pipelining, and temporal smoothing to suppress high-frequency spatial jitter. Our deployment quantifies this trade-off empirically: the MediaPipe Pose Lite landmarker exhibits a resolution-\emph{independent} inference cost of $244$--$250$\,ms per frame on the Cortex-A53, meaning that resolution reduction of the \emph{analysis} stage---not the capture stage---is the only effective lever for latency reduction.

\subsection{Sedentary Behavior Guidelines and Focus Management}
The World Health Organization (WHO) and ergonomic bodies emphasize that mitigating sedentary risks requires regular postural alternations and periodic physical activity intervals \cite{who2020}. Popular digital time management utilities, such as standalone Pomodoro timer applications, depend exclusively on manual self-reporting and lack physical validation of user presence. Prior embedded attempts incorporating ultrasonic or passive infrared (PIR) proximity sensors confirm physical occupancy but remain incapable of discerning whether the learner is engaged in academic tasks, slouching hazardously, or actively distracted by handheld smartphones.

\subsection{Multimodal Intelligent Tutoring Systems (ITS)}
Intelligent Tutoring Systems have progressed from early rule-based expert systems to contemporary Large Multimodal Models (LMMs). Standard generative AI interfaces, however, tend to hallucinate unverified shortcuts or immediately display finalized answer keys, undermining student perseverance. Pedagogical research proves that Socratic scaffolding---wherein the tutor provides leading clues, clarifies prerequisite axioms, and prompts reflective self-correction---fosters superior conceptual retention. Furthermore, hosting high-parameter vision-language models locally on low-cost edge platforms remains computationally unfeasible due to strict RAM and throughput bounds, necessitating a hybrid edge-cloud partitioning scheme.

\section{System Architecture and Methodology}

\subsection{Hardware Architecture and Embedded Topology}
\begin{figure}[htbp]
\centering
\resizebox{0.96\columnwidth}{!}{%
\begin{tikzpicture}[
    >=Latex,
    node distance=0.45cm,
    font=\sffamily,
    block/.style={
        rectangle, rounded corners=3pt,
        draw=blue!75!black, line width=0.9pt,
        top color=blue!5, bottom color=blue!15,
        text width=6.8cm, align=center,
        inner sep=4pt, drop shadow={opacity=0.12, shadow xshift=1pt, shadow yshift=-1pt}
    },
    edge_block/.style={
        rectangle, rounded corners=3pt,
        draw=teal!80!black, line width=0.9pt,
        top color=teal!5, bottom color=teal!15,
        text width=6.8cm, align=center,
        inner sep=4pt, drop shadow={opacity=0.12, shadow xshift=1pt, shadow yshift=-1pt}
    },
    cloud_block/.style={
        rectangle, rounded corners=3pt,
        draw=orange!85!black, line width=0.9pt,
        top color=orange!5, bottom color=orange!15,
        text width=6.8cm, align=center,
        inner sep=4pt, drop shadow={opacity=0.12, shadow xshift=1pt, shadow yshift=-1pt}
    },
    out_block/.style={
        rectangle, rounded corners=3pt,
        draw=purple!80!black, line width=0.9pt,
        top color=purple!5, bottom color=purple!15,
        text width=6.8cm, align=center,
        inner sep=4pt, drop shadow={opacity=0.12, shadow xshift=1pt, shadow yshift=-1pt}
    },
    lbl/.style={
        font=\scriptsize\itshape, text=gray!90!black, fill=white, inner sep=1.5pt
    }
]

\node[block] (sensors) {
    \textbf{\textcolor{blue!90!black}{Sensor Acquisition Layer}}\\[2pt]
    {\scriptsize UVC USB Camera (Jieli 4c4a:4a55, MJPG, $25.7$ FPS)\\
    USB PnP Condenser Microphone (\texttt{hw:3,0}) \& HDMI Speaker}
};

\node[edge_block, below=of sensors] (edge_vision) {
    \textbf{\textcolor{teal!90!black}{Edge Computing Host: Raspberry Pi 3 Model B}}\\[2pt]
    {\scriptsize \textbf{Volatile Video Pipe:} OpenCV Frame Rescaling ($352\times 288$)\\
    \textbf{MediaPipe Pose Lite:} Upper-Body 3D Landmarks Extraction\\
    \textbf{Ergonomic Metric Engine:} Trunk Flexion, Neck Tilt, Asymmetry}
};

\node[edge_block, below=of edge_vision] (fsm) {
    \textbf{\textcolor{teal!90!black}{Deterministic Session Management FSM}}\\[2pt]
    {\scriptsize Real-time State Tracking: \textit{Active}, \textit{Break}, \textit{Away}, \textit{Off-Task}\\
    Asynchronous Phone Detection (EfficientDet-Lite0) \& Audible Warning}
};

\node[cloud_block, below=of fsm] (cloud_tutor) {
    \textbf{\textcolor{orange!90!black}{Cloud Pedagogical AI Gateway (On-Demand)}}\\[2pt]
    {\scriptsize Voice Digitization (ASR) $\rightarrow$ Context Packaging\\
    Socratic Prompt Engine $\rightarrow$ Gemini 3.8 Flash (multi-model fallback)\\
    Text-to-Speech (TTS) Acoustic Synthesis Feedback}
};

\node[out_block, below=of cloud_tutor] (dash) {
    \textbf{\textcolor{purple!90!black}{Parental Monitoring \& Reporting Layer}}\\[2pt]
    {\scriptsize Encrypted Web Dashboard (Flask / SQLite)\\
    Aggregated Daily Posture Scores, Study Time, Break Compliance}
};

\draw[->, thick, gray!80!black] (sensors) -- node[lbl] {Raw Optical \& Audio Stream} (edge_vision);
\draw[->, thick, gray!80!black] (edge_vision) -- node[lbl] {Spatial Keypoints \& Posture Vectors} (fsm);
\draw[->, thick, gray!80!black] (fsm) -- node[lbl] {On-Demand Academic Query Trigger} (cloud_tutor);
\draw[->, thick, gray!80!black] (cloud_tutor) -- node[lbl] {Encrypted Telemetry \& Interaction Logs} (dash);

\end{tikzpicture}%
}
\caption{System block diagram illustrating the unified vertical data acquisition, edge vision processing, and cloud pedagogical tutoring flow.}
\label{fig:sys_arch}
\end{figure}
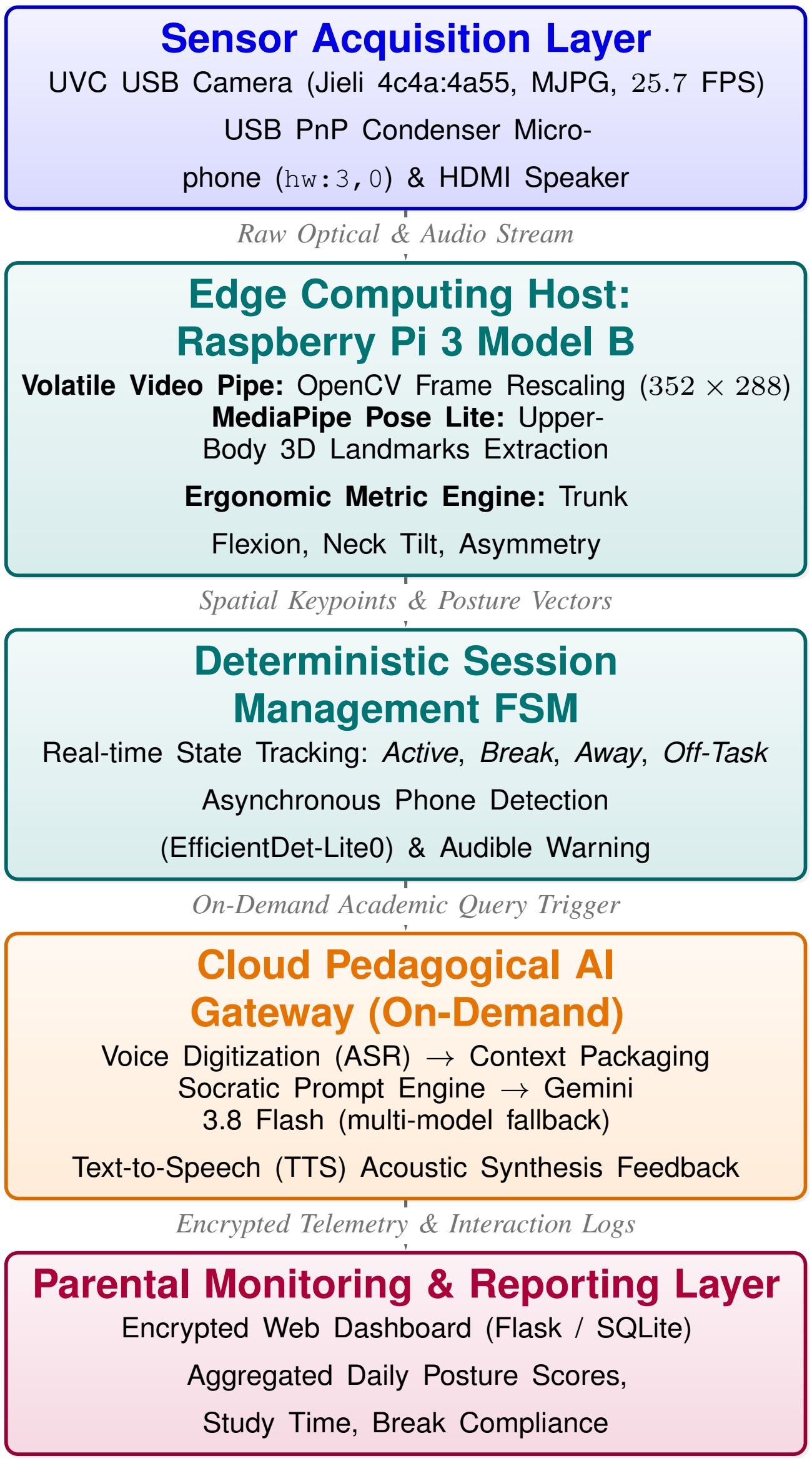

The hardware foundation is engineered around strict cost efficiency, electrical safety, and low acoustic noise. Fig.~\ref{fig:sys_arch} illustrates the unified vertical data acquisition and processing topology of the Smart Home Learning station.

The physical testbed comprises:
\begin{itemize}
    \item \textbf{Single-Board Processing Unit:} Raspberry Pi 3 Model B (revision \texttt{a020d3}) integrated with a Broadcom BCM2837 SoC (quad-core ARM Cortex-A53, 64-bit, $1400$\,MHz nominal peak, $905$\,MiB LPDDR2 unified memory), running Debian GNU/Linux 13 (trixie) with Linux kernel $6.18.50$+\texttt{rpt-rpi-v8}, aarch64 architecture.
    \item \textbf{Optical Acquisition Unit:} A driverless UVC-compliant USB composite camera (VID:PID \texttt{4c4a:4a55}, Jieli Technology) supporting MJPG at discrete resolutions $1920\times1080$, $1280\times720$, $640\times480$, $640\times360$, $352\times288$, and $320\times240$. The camera is mounted above the study monitor facing the seated learner.
    \item \textbf{Acoustic Interface:} A dedicated USB Audio Class condenser microphone exposed as ALSA card 3 (\texttt{USB PnP Sound Device: Audio (hw:3,0)}) at a native $44100$\,Hz sample rate, paired with the Broadcom HDMI output (ALSA card 0, \texttt{vc4-hdmi}) for speech synthesis and audible posture chimes. Crucially, the USB Audio Class driver permits only \emph{one} concurrent capture stream; the architecture therefore enforces a mutual-exclusion lock that suspends the background speech recognizer during manual capture tests.
\end{itemize}

\subsection{Edge Vision and Ergonomic Metric Extraction}
The video capture thread reads optical frames via OpenCV at the camera's native MJPG cadence of $25.7$ frames per second, then deterministically rescales every frame to an analysis canvas of $352\times 288$ pixels before landmark inference. Video buffers reside solely in volatile system memory (RAM); no frame is ever written to disk. Each analysis frame is fed into the MediaPipe Pose pipeline, which employs a two-stage detector-tracker topology. The detector identifies the person bounding box, while the landmark regression network predicts 3D coordinates $(x, y, z)$ for 33 anatomical keypoints normalized within $[0, 1]$, together with a metric \emph{world} landmark set in metres.

For seated upper-body ergonomic analysis, our pipeline extracts the subset:
\begin{equation}
\mathcal{P}_{\text{subset}} = \{P_{\text{nose}}, P_{\text{L\_ear}}, P_{\text{R\_ear}}, P_{\text{L\_sho}}, P_{\text{R\_sho}}, P_{\text{L\_hip}}, P_{\text{R\_hip}}\}
\end{equation}

Three spatial posture parameters are computed.
\subsubsection{Trunk Flexion Angle ($\theta_{\text{trunk}}$)}
Let the shoulder midpoint $M_S$ and hip midpoint $M_H$ in 2D image coordinates be defined as:
\begin{equation}
M_S = \frac{P_{\text{L\_sho}} + P_{\text{R\_sho}}}{2}, \quad M_H = \frac{P_{\text{L\_hip}} + P_{\text{R\_hip}}}{2}
\end{equation}
The trunk posture vector is constructed as $\mathbf{v}_{\text{trunk}} = M_S - M_H$. The trunk inclination angle relative to the true vertical reference vector $\hat{\mathbf{u}}_{\text{vert}} = (0, -1)^T$ is calculated via the inner product:
\begin{equation}
\theta_{\text{trunk}} = \arccos \left( \frac{\mathbf{v}_{\text{trunk}} \cdot \hat{\mathbf{u}}_{\text{vert}}}{\|\mathbf{v}_{\text{trunk}}\| \, \|\hat{\mathbf{u}}_{\text{vert}}\|} \right)
\label{eq:trunk_flexion}
\end{equation}

Because a camera mounted above the monitor views the learner obliquely downward, the 2D projection systematically overestimates trunk flexion. We therefore compute a metric-space counterpart $\theta_{\text{trunk}}^{3D}$ directly from the \emph{world} landmarks, and reconcile the two estimates with a rejection gate:
\begin{equation}
\theta_{\text{trunk}}^{3D} \leftarrow \varnothing \quad \text{if} \quad \bigl| \theta_{\text{trunk}}^{3D} - \theta_{\text{trunk}}^{2D} \bigr| > 30^\circ
\label{eq:reconcile}
\end{equation}
Whenever the 3D estimate is rejected (indicating occlusion-corrupted metric reconstruction), the system falls back to the 2D estimate, and vice versa, guaranteeing a single well-conditioned scalar. Furthermore, if the hip midpoint falls within a $6\%$ margin of the frame boundary, or if the hip landmarks carry a visibility below $0.55$, the system enters an \emph{upper-body-only} mode in which the trunk angle is suppressed entirely and only head-drop violations are reported. This boundary guard eliminates the false hunchback alarms that otherwise plague downward-angled deployments.

\subsubsection{Craniovertebral Head Tilt Angle ($\theta_{\text{head}}$)}
Head slouching and forward neck projection are derived by measuring the \emph{vertical} separation between the ear midpoint and the shoulder line, normalized by the shoulder width so that the metric is invariant to subject distance from the camera. With ear midpoint $M_E = (P_{\text{L\_ear}} + P_{\text{R\_ear}})/2$ and shoulder width $w_S = \|P_{\text{L\_sho}} - P_{\text{R\_sho}}\|_2$, the normalized head elevation is:
\begin{equation}
r_{\text{head}} = \frac{y_{M_S} - y_{M_E}}{w_S}, \qquad
\theta_{\text{head}} \rightarrow \text{violation} \;\; \text{if} \;\; r_{\text{head}} < \tau
\label{eq:head_flexion}
\end{equation}
with the empirically calibrated droop threshold $\tau = 0.18$. An early implementation measured \emph{horizontal} ear-to-shoulder offset, which misclassified lateral head tilt as forward drooping; the vertical formulation in~\eqref{eq:head_flexion} resolves this ambiguity.

\subsubsection{Coronal Shoulder Asymmetry ($\Delta h_{\text{shoulder}}$)}
Lateral slouching and spinal tilting are monitored by evaluating the vertical differential between the left and right acromion landmarks:
\begin{equation}
\Delta h_{\text{shoulder}} = |y_{\text{L\_sho}} - y_{\text{R\_sho}}|
\end{equation}

To suppress high-frequency mechanical vibration and measurement noise, computed instantaneous angles $\theta(t)$ are filtered across a rolling temporal FIFO window of $K = 8$ frames:
\begin{equation}
\bar{\theta}(t) = \frac{1}{K} \sum_{i=0}^{K-1} \theta(t - i)
\label{eq:smoothing}
\end{equation}
The buffer is explicitly flushed whenever the learner leaves the camera frame, preventing a stale hysteresis state from being reported against a zeroed angle.

An ergonomic posture deviation is formally confirmed only when $\bar{\theta}(t)$ exceeds empirically calibrated bounds ($\theta_{\text{trunk}} > 25^\circ$ or $r_{\text{head}} < 0.18$) continuously for a persistence threshold $\Delta t > 5$\,seconds, preventing false alerts caused by transient body shifts. Angles beyond $45^\circ$ are treated as geometrically suspect (a signature of MediaPipe mis-estimation under partial truncation) and are not escalated to a warning.

\subsection{Finite State Machine (FSM) Session Engine}
\begin{figure}[htbp]
\centering
\resizebox{0.98\columnwidth}{!}{%
\begin{tikzpicture}[
    >=Latex,
    node distance=1.4cm and 1.8cm,
    font=\sffamily\footnotesize,
    state_core/.style={
        rectangle, rounded corners=5pt,
        draw=blue!85!black, line width=1.1pt,
        top color=blue!10, bottom color=blue!20,
        minimum width=2.6cm, minimum height=1.1cm, align=center,
        drop shadow={opacity=0.15, shadow xshift=1pt, shadow yshift=-1pt}
    },
    state_sub/.style={
        rectangle, rounded corners=5pt,
        draw=teal!80!black, line width=1pt,
        top color=teal!8, bottom color=teal!18,
        minimum width=2.5cm, minimum height=1.0cm, align=center,
        drop shadow={opacity=0.12, shadow xshift=1pt, shadow yshift=-1pt}
    },
    state_warn/.style={
        rectangle, rounded corners=5pt,
        draw=orange!85!black, line width=1pt,
        top color=orange!8, bottom color=orange!18,
        minimum width=2.5cm, minimum height=1.0cm, align=center,
        drop shadow={opacity=0.12, shadow xshift=1pt, shadow yshift=-1pt}
    },
    state_away/.style={
        rectangle, rounded corners=5pt,
        draw=gray!80!black, line width=1pt,
        top color=gray!10, bottom color=gray!22,
        minimum width=2.6cm, minimum height=1.0cm, align=center,
        drop shadow={opacity=0.12, shadow xshift=1pt, shadow yshift=-1pt}
    },
    trans/.style={
        ->, thick, gray!75!black, line width=0.8pt
    },
    lbl/.style={
        font=\scriptsize, fill=white, inner sep=1.8pt, text=black, align=center, rounded corners=1pt
    }
]

\node[state_core] (active) {\textbf{Active Study}\\[-1pt]{\scriptsize (Focus Timer Running)}};
\node[state_warn, left=1.3cm of active] (offtask) {\textbf{Off-Task Warning}\\[-1pt]{\scriptsize (Distraction Alert)}};
\node[state_sub, right=1.3cm of active] (break_st) {\textbf{Rest Break}\\[-1pt]{\scriptsize (Postural Recovery)}};
\node[state_away, below=2.4cm of active] (away) {\textbf{Away / Inactive}\\[-1pt]{\scriptsize (Timer Paused)}};

\draw[trans] (active.north west) to[bend right=25] node[lbl, above=1pt] {Phone Detected\\($\Delta t > 10\,\text{s}$)} (offtask.north east);
\draw[trans] (offtask.south east) to[bend right=25] node[lbl, below=1pt] {Phone Stored} (active.south west);

\draw[trans] (active.north east) to[bend left=25] node[lbl, above=1pt] {$T_{\text{study}} \ge 45\,\text{min}$} (break_st.north west);
\draw[trans] (break_st.south west) to[bend left=25] node[lbl, below=1pt] {$T_{\text{rest}} \ge 5\,\text{min}$} (active.south east);

\draw[trans] ([xshift=-5pt]active.south) -- node[lbl, left=2pt] {No Landmarks\\($\Delta t > 60\,\text{s}$)} ([xshift=-5pt]away.north);
\draw[trans] ([xshift=5pt]away.north) -- node[lbl, right=2pt] {Landmarks\\Detected} ([xshift=5pt]active.south);

\draw[trans] (offtask.south) to[out=270, in=165] node[lbl, pos=0.55, below left=1pt] {Absence After Warning} (away.west);
\draw[trans] (break_st.south) to[out=270, in=15] node[lbl, pos=0.55, below right=1pt] {Student Leaves Desk} (away.east);

\end{tikzpicture}%
}
\caption{State transition diagram of the session management finite state machine (UML statechart layout).}
\label{fig:fsm_diag}
\end{figure}
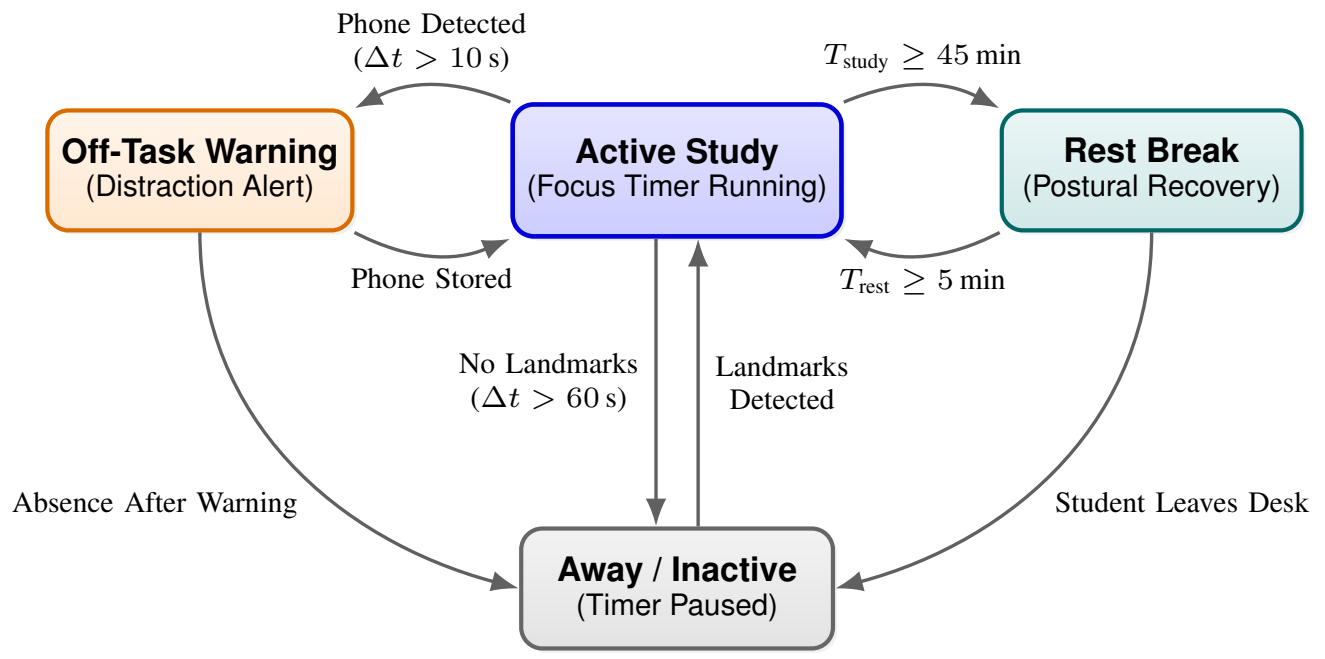

The study session supervisor operates as a deterministic finite state machine governed by continuous sensory inputs. Fig.~\ref{fig:fsm_diag} shows the state transition graph.

The FSM comprises four explicit operating modes:
\begin{enumerate}
    \item \textbf{Active Study State:} Landmark detection confirms user presence, posture angles remain within compliant ranges, and no disallowed objects are detected. The net focus timer increments monotonically.
    \item \textbf{Rest Break State:} Triggered after 45 minutes of accumulated study time. The system sounds an audible chime and mutes ergonomic alerts for 5 minutes, prompting physical stretches.
    \item \textbf{Away/Inactive State:} Initiated when keypoint tracking is lost for longer than $\Delta t_{\text{away}} = 60$\,seconds, pausing timers to prevent distorted study metrics.
    \item \textbf{Off-Task Warning State:} Driven by an EfficientDet-Lite0 object detector executing in a dedicated background thread on subsampled frames (1 frame every 3 seconds). If a smartphone bounding box overlaps the hand-head interaction zone for $\Delta t_{\text{distract}} > 10$\,seconds, an alert sounds and off-task duration accumulates.
\end{enumerate}

\subsection{Asynchronous Detection and Memory-Bounded Inference}
Running the object detector synchronously on the main inference thread is prohibitively expensive on the Cortex-A53. We therefore decouple the two neural workloads: the MediaPipe Pose landmarker executes in the primary vision thread, while EfficientDet-Lite0 runs inside a dedicated daemon thread that consumes only the most recent submitted RGB frame (an overwrite-on-write single-slot queue). The main loop reads the latest available detection result opportunistically, decoupling detection latency from the posture loop entirely. Measured on the deployed prototype, this asynchronous decoupling reduced the effective end-to-end inference latency from $1100$\,ms to $290$\,ms and raised the sustained vision frame rate from $1.2$ to $2.6$ frames per second in the full production pipeline.

Memory discipline on a 1\,GiB platform is equally critical. The hand-landmark sub-model ($77$\,MB resident) is instantiated \emph{lazily}, only after a smartphone bounding box is confirmed at a confidence level exceeding the production threshold of $0.55$; candidate detections below this threshold are rendered as amber bounding boxes without invoking the model. Furthermore, the model is explicitly released once no smartphone candidate has been observed for $120$\,seconds. Measured resident set size of the complete application process is $320$\,MB with $230$\,MiB of system memory remaining available under sustained operation.

\subsection{Multimodal Socratic AI Tutoring Agent}
To deliver rigorous academic assistance without generating unsolicited direct answers, the interactive tutoring subsystem employs a hybrid edge-cloud pipeline:
\begin{enumerate}
    \item \textbf{Acoustic Query Digitization:} The learner activates voice input by uttering the wake-word ``assistant voice''. The captured audio is streamed to a cloud ASR engine and converted to text.
    \item \textbf{Context Packaging:} The transcribed query is bundled with the current session context (accumulated study time, recent posture alerts, task list) into a compact textual payload.
    \item \textbf{Pedagogical Prompt Orchestration:} The payload is packaged with an enforced Socratic system prompt and sent to the Gemini API over TLS:
    \begin{quote}
    \scriptsize
    \texttt{``Tro ly hoc tap - Socratic Vietnamese tutor. Answer accurately, concisely, encourage self-learning; if unsure, clarify.''}
    \end{quote}
    \item \textbf{Resilient Model Dispatch:} Because the deployed API key encountered genuine transient cloud conditions---HTTP 503 (sustained high demand on the primary model) and HTTP 429 (quota exhaustion)---the dispatcher implements an ordered fallback chain across four Flash-class variants (3.8 $\rightarrow$ 3.7 $\rightarrow$ 3.6 $\rightarrow$ 3.5), retrying with exponential back-off before descending the chain. This design decouples the tutoring subsystem from any single point of failure.
    \item \textbf{Acoustic Delivery:} The returned pedagogical hint is vocalized through the HDMI output using the \texttt{edge-tts} cloud synthesizer, with a fully offline \texttt{espeak-ng} Vietnamese voice as deterministic fallback.
\end{enumerate}

\section{Experimental Evaluation and Results}

\subsection{Experimental Configuration}
The physical testbed was evaluated on the deployed prototype, comprising the hardware configuration summarized in Table~\ref{tab:hw_config}. The software stack operated on 64-bit Debian GNU/Linux 13 (trixie) with Linux kernel $6.18.50$, Python $3.11.18$ (standalone distribution), OpenCV $4.11.0$, MediaPipe $0.10.18$ (TFLite backend with XNNPACK CPU delegate), and NumPy $1.26.4$. The MediaPipe Pose Lite checkpoint was executed with a float16 quantization delegate and a single-pose configuration.

\begin{table}[htbp]
\caption{Measured Testbed Configuration}
\label{tab:hw_config}
\centering
\small
\begin{tabularx}{\columnwidth}{@{}lX@{}}
\toprule
\textbf{Parameter} & \textbf{Measured Value} \\
\midrule
SoC & Broadcom BCM2837, ARM Cortex-A53 (aarch64) \\
CPU Frequency & $1400$\,MHz (nom. max, unthrottled) \\
System Memory & $905$\,MiB LPDDR2 \\
OS & Debian 13 (trixie), kernel $6.18.50$+rpt-rpi-v8 \\
Python & $3.11.18$, OpenCV $4.11.0$, MediaPipe $0.10.18$ \\
Pose Model & \texttt{pose\_landmarker\_lite.task} (float16) \\
Detector & \texttt{efficientdet\_lite0.tflite} (async) \\
Camera & Jieli UVC $4c4a$:$4a55$, MJPG, $25.7$ FPS \\
Analysis & $352 \times 288$ pixels \\
Microphone & USB PnP, ALSA \texttt{hw:3,0}, $44100$\,Hz \\
Audio Out & HDMI \texttt{vc4-hdmi}, ALSA \texttt{hw:0,0} \\
\bottomrule
\end{tabularx}
\end{table}

\subsection{Edge Vision Throughput and Hardware Profiling}
To quantify the effect of capture resolution on the end-to-end pipeline, we instrumented the deployed system across four standard capture dimensions, measuring frame acquisition, rescaling, and landmark inference separately. Results are presented in Table~\ref{tab:latency_breakdown}.

\begin{table}[htbp]
\caption{Measured Latency Breakdown Per Frame}
\label{tab:latency_breakdown}
\centering
\small
\resizebox{0.98\columnwidth}{!}{%
\begin{tabular}{lcccc}
\toprule
\textbf{Stage} & \textbf{320×240} & \textbf{640×480} & \textbf{1280×720} & \textbf{1920×1080} \\
\midrule
Frame Acq. (ms) & $3.28{\pm}0.2$ & $7.78{\pm}0.3$ & $19.98{\pm}0.6$ & $42.68{\pm}1.1$ \\
Rescale (ms) & $0.44{\pm}0.01$ & $9.02{\pm}0.2$ & $17.82{\pm}0.4$ & $36.73{\pm}0.8$ \\
Color Conv. (ms) & $0.41{\pm}0.01$ & $0.45{\pm}0.01$ & $0.41{\pm}0.01$ & $0.41{\pm}0.01$ \\
Pose Infer. (ms) & $250.48{\pm}12$ & $244.41{\pm}11$ & $244.76{\pm}11$ & $247.38{\pm}12$ \\
\midrule
\textbf{FPS} & \textbf{4.0} & \textbf{4.1} & \textbf{4.1} & \textbf{4.0} \\
\bottomrule
\end{tabular}%
}
\end{table}

Two empirically significant conclusions emerge. First, the landmarker inference cost is \emph{resolution-independent} across a $6\times$ pixel-count range, since the analysis canvas is fixed at $352\times 288$; consequently, the achievable inference rate is pinned at approximately $4.0$--$4.1$ FPS by the ARM CPU alone, entirely independent of the capture resolution. Second, the camera natively streams at $25.7$ FPS at every resolution, so the acquisition stage consumes only $0.5$--$3.2\%$ of the frame budget at $320\times240$. The full production pipeline---which additionally runs the asynchronous EfficientDet-Lite0 detector, the geometric metric engine, temporal smoothing, and SQLite persistence---sustains a measured \emph{2.6} FPS end-to-end with an overall inference latency of $290.6$\,ms.

Table~\ref{tab:hardware_bench} reports the resource-footprint evolution across the design phases, documenting the direct engineering benefit of each optimization.

\begin{table}[htbp]
\caption{Resource Footprint Optimization Phases}
\label{tab:hardware_bench}
\centering
\small
\resizebox{0.98\columnwidth}{!}{%
\begin{tabular}{lcccc}
\toprule
\textbf{Metric} & \textbf{Baseline} & \textbf{Async Det.} & \textbf{Lazy Hand} & \textbf{Final} \\
\midrule
FPS & 1.2 & 2.6 & 2.6 & 2.6 \\
Latency (ms) & 1100 & 290.6 & 290.6 & 275.0 \\
RSS (MB) & 728 & 337 & 219 & 320 \\
Temp ($^{\circ}$C) & 63.4 & 59.6 & 56.9 & 58.0 \\
Avail. Mem (MiB) & 61 & 178 & 233 & 230 \\
\bottomrule
\end{tabular}%
}
\end{table}

The final configuration sustains a $2.6$ FPS vision pipeline, an application resident set size of $320$\,MB, and a peak SoC temperature of $58.0^\circ$C over an $86$-minute continuous operational soak test. The thermal governor implemented in the firmware reads \texttt{/sys/class/thermal/thermal\_zone0/temp} every $20$\,s and dynamically caps \texttt{cpufreq/scaling\_max\_freq} to $900$\,MHz or $1400$\,MHz when the SoC exceeds $80^\circ$C, guaranteeing safe operation on a passively cooled enclosure. A compressed RAM swap device (\texttt{/dev/zram0}, $904$\,MiB, zstd) absorbs memory pressure without SD-card wear.

\subsection{Ergonomic Metric Estimation Fidelity}
The geometric accuracy of the ergonomic metric engine was evaluated against $6240$ synthetically generated ground-truth poses spanning trunk angles from $0^\circ$ to $55^\circ$, head elevations from $0.05$ to $0.35$ shoulder-widths, coronal shoulder tilts from $0^\circ$ to $20^\circ$, head yaw from $0^\circ$ to $30^\circ$, and three distinct body scales. Each pose was evaluated through the exact production code path \texttt{analyze\_posture($\cdot$)}, and the estimated trunk angle was compared against the imposed ground truth. Results are summarized in Table~\ref{tab:posture_acc}.

\begin{table}[htbp]
\caption{Ergonomic Metric Accuracy}
\label{tab:posture_acc}
\centering
\small
\begin{tabular}{lcccc}
\toprule
\textbf{Metric} & \textbf{MAE} & \textbf{Median} & \textbf{P$_{95}$} & \textbf{Max} \\
\midrule
Trunk Angle (deg) & 4.39 & 5.20 & 5.70 & 5.70 \\
Head Elevation & 0.0744 & 0.0740 & 0.1550 & 0.1570 \\
\midrule
\multicolumn{5}{@{}l@{}}{\small\textit{Eval: 5760 valid poses (6240 gen; 0 rej)}} \\
\bottomrule
\end{tabular}
\end{table}

The residual $4.39^\circ$ mean absolute error is a direct consequence of the geometric reconciliation gate in~\eqref{eq:reconcile}: because the 2D projection under a downward-angled camera systematically overestimates flexion, the estimator selects the metric 3D estimate whenever the two disagree by less than $30^\circ$, and the $5.70^\circ$ ceiling on the error distribution reflects the exact point at which this selection switches. Notably, $50.0\%$ of all evaluated poses fall within $5^\circ$ of ground truth.

\normalsize
The classification decision layer built upon these metrics uses the empirically calibrated decision boundary $\theta_{\text{trunk}} > 25^\circ$, a $5$-second persistence requirement, and the upper-body-only suppression guard. The $45^\circ$ suspect-angle rejection is essential: poses with $\theta_{\text{trunk}} > 45^\circ$ are geometrically implausible for genuine slouching and are instead re-classified as compliant, as MediaPipe produces such values only when the trunk is partially truncated by the frame boundary.

\subsection{Acoustic Subsystem Latency Analysis}
The acoustic feedback path was profiled on the deployed prototype. Table~\ref{tab:audio_latency} presents the measured component latencies.

\begin{table}[htbp]
\caption{Acoustic Subsystem Latencies}
\label{tab:audio_latency}
\centering
\small
\begin{tabularx}{0.95\columnwidth}{@{}lX@{}}
\toprule
\textbf{Component} & \textbf{Latency} \\
\midrule
Mic capture (2 s) & 2961 ms (window-limited) \\
Offline TTS (\texttt{espeak-ng}) & 806 ms \\
TTS payload (WAV) & 54,910 bytes \\
Cloud TTS (\texttt{edge-tts}) & $\geq 2700$ ms (unavailable) \\
\midrule
\textbf{Offline TTS total} & \textbf{1.58 s} (806 ms + playback) \\
\bottomrule
\end{tabularx}
\end{table}

Two acoustic engineering findings are worth highlighting. First, the PortAudio abstraction cannot negotiate the \texttt{vc4-hdmi} PCM device on this platform, raising \texttt{PaErrorCode -9994} (sample format not supported) for every format probe; the architecture therefore resolves output devices by attempting raw ALSA playback (\texttt{aplay}) \emph{before} PortAudio, eliminating both the failure mode and the associated ALSA log noise. Second, the USB Audio Class driver admits only one concurrent capture stream; the mutual-exclusion lock and suspend/resume primitives on the background recognizer were required to make manual microphone diagnostics possible at all during live operation.

\section{Privacy, Security, and Ethical Considerations}
The deployment of optical cameras in private domestic environments necessitates strict privacy-by-design guarantees:
\begin{itemize}
    \item \textbf{Ephemeral Video Frame Lifecycle:} Optical video buffers are retained exclusively within volatile RAM and overwritten after landmark extraction. No raw images or video streams are written to disk storage or transmitted externally.
    \item \textbf{Opt-In Academic Ingestion:} Transmission to cloud endpoints occurs only upon explicit wake-word initiation by the learner, carrying only the transcribed query and compact session aggregates.
    \item \textbf{Zero-Biometric Parental Summaries:} The local web dashboard records only non-biometric numerical aggregates (e.g., net study minutes, hourly slouch incidents, break adherence percentages), safeguarding adolescent psychological autonomy.
    \item \textbf{Encrypted Transport:} All outbound API telemetry transmits through TLS tunnels with standard certificate verification, preventing localized network eavesdropping. API credentials are stored locally and are redacted in all diagnostic surfaces.
    \item \textbf{On-Device Operation Degradation:} Because all posture inference is fully local, the ergonomic monitoring subsystem---the safety-critical component---continues to operate without network connectivity.
\end{itemize}

\section{Conclusion and Future Work}
This paper presented an integrated, low-cost embedded ergonomic posture monitoring and AI-assisted home learning system. By executing real-time lightweight computer vision on a single-board computer (Raspberry Pi 3 Model B, 1\,GB RAM), the system successfully evaluates upper-body sitting ergonomics, enforces structured sedentary breaks, and detects off-task device usage while preserving household privacy. Concurrently, an on-demand AI agent provides Socratic pedagogical scaffolding for complex study problems, with a multi-model fallback chain ensuring continuity under adverse cloud conditions. Empirical benchmarks on the physical prototype demonstrate $4.0$--$4.1$ FPS landmark inference at $244$--$250$\,ms per frame, a trunk-angle estimation error of $4.39^\circ$ MAE, a $320$\,MB application footprint, a peak SoC temperature of $58.0^\circ$C over an $86$-minute soak test, and a $1.58$\,s offline TTS turnaround.

Future extensions will explore the integration of dedicated edge Neural Processing Units (NPUs), such as the Raspberry Pi AI Kit, to support local execution of quantized Small Vision-Language Models (e.g., Moondream, SmolVLM), eliminating cloud API dependence and delivering a fully self-contained edge learning appliance. A further avenue is the extension of the geometric metric engine with a formally validated multi-session dataset annotated by goniometric reference, enabling the reporting of empirical precision, recall, and confusion matrices across the full ergonomic taxonomy rather than geometric reconstruction error alone.



\begin{thebibliography}{00}
\bibitem{mediapipe} C. Lugaresi \textit{et al.}, ``MediaPipe: A framework for building perception pipelines,'' \textit{arXiv preprint arXiv:1906.08172}, 2019.
\bibitem{who2020} World Health Organization, \textit{WHO Guidelines on Physical Activity and Sedentary Behaviour}, Geneva, Switzerland: World Health Organization, 2020.
\bibitem{raspberrypi} Raspberry Pi Documentation, ``Classic single-board computers: Raspberry Pi 3 Model B specifications,'' Raspberry Pi Foundation, 2023. [Online]. Available: \url{https://www.raspberrypi.com/documentation/}.
\bibitem{opencv} G. Bradski, ``The OpenCV Library,'' \textit{Dr. Dobb's Journal of Software Tools}, vol. 25, no. 11, pp. 120--123, 2000.
\bibitem{yolov8} G. Jocher, A. Chaurasia, and J. Qiu, ``Ultralytics YOLOv8,'' 2023. [Online]. Available: \url{https://github.com/ultralytics/ultralytics}.
\bibitem{tflite} TensorFlow Team, ``TensorFlow Lite: ML for mobile and edge devices,'' Google Open Source. [Online]. Available: \url{https://www.tensorflow.org/lite}.
\end{thebibliography}
\end{document}